\documentclass[pra,aps,twocolumn,amsmath,amssymb]{revtex4}
\usepackage{graphicx}
\usepackage{dcolumn}
\usepackage{bm}
\usepackage[extension=xxx]{hyperref}

\newcommand{\beq}{\begin{equation}}
\newcommand{\eeq}{\end{equation}}
\newcommand{\beqa}{\begin{eqnarray}}
\newcommand{\eeqa}{\end{eqnarray}}

\def\beq{\begin{equation}}

\usepackage{color}
\begin{document}

\title{Transport in a harmonic trap: shortcuts to adiabaticity and robust protocols}
\date{\today}

\author{D. Gu\'ery-Odelin$^{1,2}$}
\author{J. G. Muga$^{3,4}$}

\affiliation{$^{1}$Universit\'e de Toulouse ; UPS ; Laboratoire Collisions Agr\'egats R\'eactivit\'e, IRSAMC ; F-31062 Toulouse, France} 
\affiliation{$^{2}$CNRS ; UMR 5589 ; F-31062 Toulouse, France}
\affiliation{$^{3}$Departamento de Qu\'{\i}mica F\'{\i}sica, Universidad del Pa\'{\i}s Vasco - Euskal Herriko Unibertsitatea, 
Apdo. 644, 48080 Bilbao, Spain}
\affiliation{$^{4}$Department of Physics, Shanghai University, 200444 Shanghai, People's Republic of China}
%

\begin{abstract}
We study the fast  transport of a particle or a Bose-Einstein condensate  in a harmonic potential.   
An exact expression for the final excitation energy in terms of the Fourier transform of the trap
acceleration is used to engineer optimal transport protocols (with no final excitation) that are robust with respect to spring-constant errors.
The same technique provides a way to design the simultaneous and robust transport of a few non-interacting species that experience
different harmonic trap frequencies in the same trap.
\end{abstract}
\maketitle
\section{Introduction}
The accurate control of atomic motion is one of the  
goals of modern atomic, molecular, and optical (AMO) Physics. 
Neutral atoms and ions, individually, in condensates or in thermal clouds, are moved in many experimental settings 
from preparation to science chambers, to implement interferometers and 
metrological devices, or for quantum information operations among processing and storing sites.    
Atoms may simply be launched into free-flight or free-fall orbits, or be driven along moving traps. 
Moving traps offer the advantage of a more precise control that would let the atom, for example, start at rest and finish   
translationally unexcited, be stopped or accelerated, and perform curved trajectories in complex circuit geometries. 
Controlled atomic motion is a requisite for current and potential quantum technologies in which preserving quantum coherence 
and achieving high final fidelities with respect to target states is of paramount importance.  Since the effects of noise 
and decoherence increase with process time, several groups have recently developed 
fast and {robust} transport protocols, shortcuts to adiabaticity (STA) \cite{reviewSTA13},  
which are generically non-adiabatic but reproduce the final 
populations of an adiabatic transport process. 
They not only minimize the effects of noise but also allow for 
quick information processing or for more repetitions of the operation  \cite{NIST06,David08,Calarco09,Masuda10,Erik11,transOCT12,Erikcond12,Masuda12,Mikel13,Adol14,Uli14,Lu14}. 
Several experiments have also demonstrated STA transport for different systems and conditions \cite{David08,Nice,Bowler,Walther}. 

The ``robustness" mentioned before is a relative concept that depends on the type of perturbation
that affects the ideal external driving \cite{Andreas,AndreasPRL,Lu14}. One of the basic problems that protocol design must face is the instability of the 
spring constant  from run to run of the experiment, while keeping a constant value during each run. 
This may in fact be the dominant problem in some settings \cite{Lu14}.
In this paper we address this difficulty by providing a protocol design strategy that imposes zeros of final excitation in a discrete set of spring constants. With an appropriate arrangement of points, broad trap-frequency windows for excitation-free fast transport may be achieved.         

In Sec.~\ref{one body} and Appendices the expression of the excitation energy for the transport of one atom in a harmonic trap is  worked out. 
Sections \ref{two} and \ref{urp} develop the 
robust protocols, and the final discussion in Sec.~\ref{dis} points out further applications of the method.  
\section{Transport of a one-body wave function\label{one body}}
The transport of a wave function of a particle of mass $m$ 
in a moving harmonic trap of angular frequency $\omega_0$ (spring constant $m\omega_0^2$) is
described, in an effectively one dimensional  scenario, 
by the time-dependent Schr\"odinger equation
\begin{equation}
i\hbar \frac{\partial \Psi(x,t)}{\partial t} = H_0(x,t)\Psi(x,t),
\label{eq2s}
\end{equation}
where $H_0(x,t) = p^2/2m + V(x,t)$ and  $V(x,t)=m\omega_0^2 [x-x_0(t)]^2/2.$ The scalar ``transport function'' $x_0(t)$ denotes the position of the minimum of the harmonic potential and obeys the boundary conditions $x_0(0)=0$, $\dot x_0(0)=0$, $x_0(t_f)=d$, and $\dot x_0(t_f)=0$ for a transport over a distance $d$ in a time interval $t_f$, assuming that the trap starts and ends up at rest. In Appendix A we give the expression for the 
exact solution of Eq.~(\ref{eq2s}) whatever is $x_0(t)$. This solution is related to the evolution of a fictitious
classical particle of coordinate $x_c(t)$ that obeys the equation of motion of the classical counterpart of the transport problem,
\begin{equation}
\ddot x_c + \omega_0^2(x_c-x_0)=0.
\label{transp}
\end{equation}
When the initial wave function is in the ground state, the energy at a time $t$ can be readily calculated from the solution $\Psi(x,t)$ (see Appendix B):
\begin{eqnarray}
E(t) - E(0)  & = &  \langle \Psi (t) | H_0(t) | \Psi (t) \rangle -\langle \Psi (0) | H_0(0) | \Psi (0) \rangle
\nonumber \\
& =  & \frac{1}{2}m \dot x_c^2 + \frac{1}{2}m \omega_0^2 (x_c-x_0)^2.
\end{eqnarray}
If the goal is to avoid any excess energy, $E(t_f)=E(0)$, the fictitious particle trajectory must obey the boundary conditions $x_c(0)=x_0(0)=0$, $x_c(t_f)=x_0(t_f)=d$, $\dot x_c(0)=\dot x_c(t_f)=0$. By virtue of the equation fulfilled by $x_c$, 
we have also $\ddot x_c(0)=\ddot x_c(t_f)=0$, as $x_0(t)$ is assumed to be continuous. 
This set of boundary conditions coincides with that obtained with the shortcut-to-adiabaticity method based on Lewis-Riesenfeld invariants \cite{Erik11,reviewSTA13}. The solution for the transport function is then constructed by an inverse-engineering method which consists in ({\it i}) choosing a function $x_c^{\rm STA}$ that obeys the appropriate set of boundary conditions, and ({\it ii}) then inferring from Eq.~(\ref{transp}) the corresponding trajectory of the potential, $x_0^{\rm STA}$.

As shown in \cite{Erikcond12} the wave function of a Bose-Einstein condensate satisfying a Gross-Pitaevskii equation in a moving external harmonic potential is shape invariant and the only
possible excitations associated with such a mode are center of mass oscillations, along the classical trajectory
in Eq. (\ref{transp}), with constant
mean field energy. The same STA strategy as for the one-body wave function can therefore be used for the transport of a condensate.   
\section{Transport without final excess energy at two trap frequencies\label{two}}
The optimal transport of a particle corresponds to the cancellation of the Fourier transform
of the acceleration of the displacement of the trap, $\ddot x_0$, at the trap 
angular frequency $\omega_0$ (see Appendix B):
\begin{equation}
{\cal V}(\omega_0)=\bigg|  \int_0^{t_f} \ddot x_0(t') e^{-i\omega_0t'}dt'\bigg| =0.
\end{equation}
Consider two angular frequencies $\omega_1$ and $\omega_2$ 
for which we would like the final state to have no excitation: ${\cal V}(\omega_1)={\cal V}(\omega_2)=0$.  
To design the appropriate acceleration function, $\ddot x_0(t)$, we introduce an auxiliary function $g(t)$ such that 
\begin{equation}
\ddot x_0(t)=\frac{{\rm d}^4g}{{\rm d}t^4} + (\omega_1^2+\omega_2^2)\frac{{\rm d}^2g}{{\rm d}t^2} + \omega_1^2 \omega_2^2g(t),
\label{eq2}
\end{equation}
and which obeys the boundary conditions 
$g(t_f)=g(0)=g'(t_f)=g'(0)=g''(t_f)=g''(0)=g^{(3)}(x_f)=g^{(3)}(0)=0$. 
The auxiliary function is defined through the differential equation (\ref{eq2}) so that, after integrating by parts 
and taking into account these boundary conditions, the   Fourier transform of the acceleration becomes the product of a polynomial in $\omega^2$ with the desired zeros by the Fourier transform of the auxiliary function:  
\begin{equation}
{\cal V}(\omega)=\bigg| (\omega^2-\omega_1^2)(\omega^2-\omega_2^2)  \int_{0}^{t_f} e^{-i\omega t'}g(t') {\rm d}t'\bigg|.
\end{equation}
The squared frequencies in $(\omega^2-\omega_1^2)(\omega^2-\omega_2^2)$ 
imply zeros at positive and negative frequencies. The latter 
might seem to be superfluous, but they        
avoid imaginary factors  in Eq.~(\ref{eq2}) and guarantee the reality of $x_0(t)$
for real $g$.    

After designing $g(t)$ and deducing $\ddot x_0$ via Eq.~(\ref{eq2}), we integrate this equation twice with the boundary conditions  $x_0(0)=0$, $\dot x_0(0)=0$, $x_0(t_f)=d$ and $\dot x_0(t_f)=0$ to specify the transport function,  
\begin{equation}
x_0(t) =\int_0^t {\rm d}t'\int_0^{t'}{\rm d}t''\ddot x_0(t'').
\end{equation}
Those latter boundary conditions imply that
\begin{equation}
\int_0^{t_f}g(t){\rm d}t=0\;{\rm and}\;
\int_0^{t_f}{\rm d}t'\int_0^{t'}g(t''){\rm d}t''=\frac{d}{\omega_1^2\omega_2^2}.
\label{eqnbc}
\end{equation}
Consider for instance the following polynomial interpolation:
\beq\label{gpol}
g(t)={\cal N}(t/t_f)^4(1-t/t_f)^4(1-2t/t_f).
\eeq
The normalization factor ${\cal N}$ is deduced from the second condition in Eq.~(\ref{eqnbc}), 
\beq\label{nor}
{\cal N}=d/(\omega_1^2\omega_2^2 t_f^2 \Delta ),
\eeq 
with 
\beqa
\Delta&=&-{\rm B}_{0}(5,5)+3{\rm B}_{0}(6,5)-2{\rm B}_{0}(7,5)
\nonumber\\
&+&{\rm B}_{1}(5,5)-3{\rm B}_{1}(6,5)+2{\rm B}_{1}(7,5),
\eeqa
where $B_z(u,v)$ is the incomplete beta Euler function. The second and third factors in Eq.~(\ref{gpol}) guarantee
the boundary conditions at the time edges 
and the fourth one provides the odd symmetry to satisfy the first integral condition in Eq.~(\ref{eqnbc}). We therefore obtain an exact analytical solution for the transport problem which fulfills exactly the desired boundary conditions.  

Finally, note that it is possible to set $\omega_1=\omega_2$. The effect is to increase (double) the multiplicity of the zero at 
$\omega=\omega_1$ which flattens the excitation energy at that point. Examples to illustrate this effect and its applications  
are worked out in the following section.     
\section{Robust protocols\label{urp}}
%
%
%
%
%
\begin{figure}[h]
\includegraphics[width=8cm]{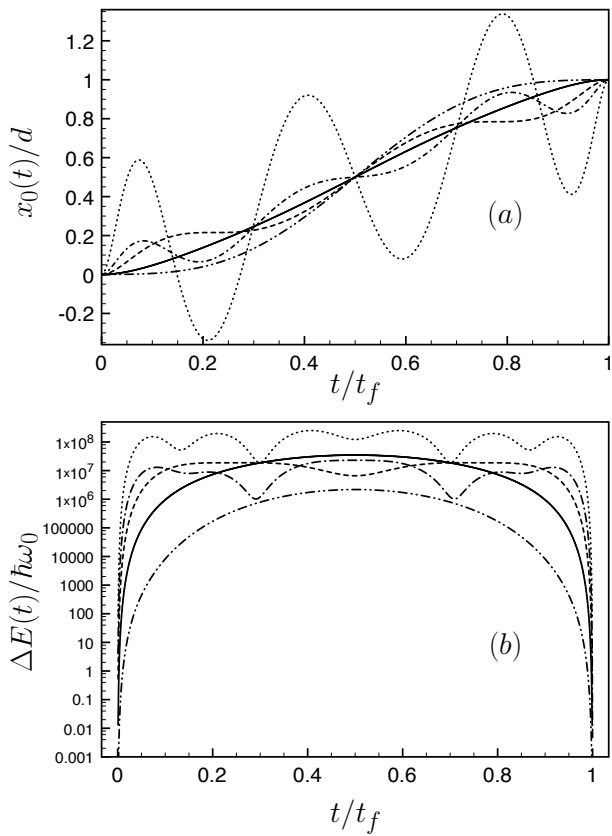}
\caption{Transport of a particle by moving its harmonic confinement of angular frequency $\omega_0$. The  parameters are chosen for the transport of $^{40}$Ca$^+$ ions over a distance of 0.4 $\mu$m (see text). (a) The trajectory of the bottom of the trap, $x_0(t)/d$ for a one (solid line, $\omega_1=\omega_0$), two (dashed line, $\omega_1=\omega_2=\omega_0$) and three (dotted line, $\omega_1=\omega_2=\omega_3=\omega_0$) frequency robust protocol for a final time $t_f$ given by $\omega_0t_f=2\pi\times 1.25$. The results of the one-point protocol with $omega_0 t_f = 2\pi \times 5$ are represented as a double dot-dashed curve. The dotted-dashed line corresponds to the same kind of 3-point protocol but for a transport duration increased by 25 \%: $\omega_0t_f=1.5625\times 2\pi$. This slight increase of the transport duration reduces dramatically the amplitude of the trap center trajectory.  (b) Variation in log-scale of the transient excess energy (in units of the quantum of energy $\hbar\omega_0$).}
\label{fig1}
\end{figure}
%
%
Robust protocols can be designed by generalizing the previous idea. Suppose that we identify an angular frequency region $[\omega_0(1-\eta),\omega_0(1+\eta)]$ 
in which the values of the trap frequencies corresponding to different runs of the experiment are distributed. 
We would like our transport protocol to provide 
excitation-free final states in this region.  
For this purpose, we choose $N$ angular frequencies $\{ \omega_1<\omega_2, ..., <\omega_N \}$, with $\omega_1<\omega_0<\omega_N$, 
and $\omega_N-\omega_1\approx 2\omega_0\eta$. 
The function $g(t)$ should have now $4N$ vanishing boundary conditions: 
\beqa
&&g(0)=g(t_f)=0, g^{(1)}(0)=g^{(1)}(t_f)=0, 
\nonumber\\
&&..., g^{(2N-1)}(0)=g^{(2N-1)}(t_f)=0,
\label{bcg}
\eeqa
where $g^{(k)}\equiv\frac{{\rm d}^kg}{{\rm d}t^k}$. 
For instance, we can use the following simple polynomial interpolation,  $g(t)={\cal N}(t/t_f)^{2N}(1-t/t_f)^{2N}(1-2t/t_f)$. The normalization factor ${\cal N}$ is determined in the same manner as previously, see Eq. (\ref{nor}), 
with 
\beqa
\!\!\!\!\!\!\!\!\!\!\Delta\!&=&\!-{\rm B}_{0}(1+2N,1+2N)+3{\rm B}_{0}(2+2N,1+2N)
\nonumber\\
&-&\!\!2{\rm B}_{0}(3+2N,1+2N)
+{\rm B}_{1}(1+2N,1+2N)
\nonumber\\
&-&\!\!3{\rm B}_{1}(2+2N,1+2N)+2{\rm B}_{1}(3+2N,1+2N).
\eeqa
%

\begin{figure}[h]
\includegraphics[width=8cm]{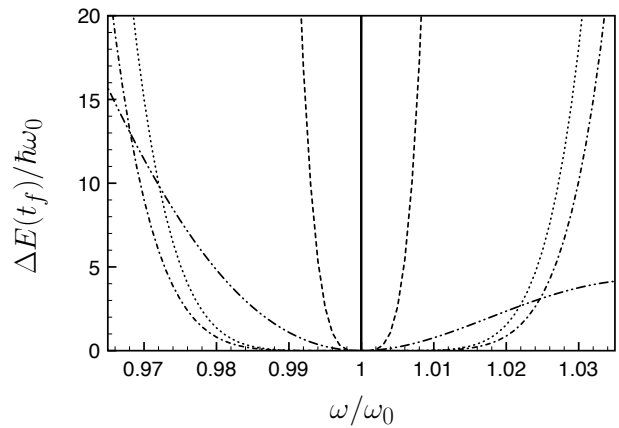}
\caption{Excess of energy $\Delta E(t_f)$ acquired after the transport normalized to the quantum of energy $\hbar\omega_0$ when the trap 
angular frequency $\omega$ differs from the optimal choice $\omega_0$. Same notations as in Fig.~\ref{fig1}.}
\label{fig1b}
\end{figure}
The desired form of the Fourier transform is
\begin{equation}
{\cal V}(\omega)= \bigg|\left( \prod_{i=1}^N(\omega_i^2-\omega^2)\right) \left( \int_{0}^{t_f} e^{-i\omega t'}g(t') {\rm d}t'\right)\bigg|.
\end{equation}
As 
\beq
 \prod_{i=1}^N(\omega_i^2-\omega^2)=\sum_{j=0}^N P_j (-1)^j ({\omega^2})^{N-j}, 
\end{equation}
where 
\beqa
P_0&=&1, P_1=\sum_{i} \omega_i^2, P_2=\sum_{i<j} \omega_i^2\omega_j^2,
\nonumber\\
P_3&=&\!\!\!\sum_{i<j<k}\!\omega_i^2\omega_j^2\omega_k^2, ..., P_N= \omega_1^2\omega_2^2...\omega_N^2,
\eeqa
we can infer the link between the acceleration $\ddot x_0$ and the auxiliary function $g$:
\begin{equation}
\ddot x_0 = 
P_0 g^{(2N)} +P_1g^{(2N-2)}+...
+P_jg^{(2N-2j)}+...+ P_Ng(t),
\end{equation}
where we have used that   
\beq
\int_0^{t_f} e^{-i\omega t'}g^{(k)}(t')=(-i\omega)^k \int_0^{t_f} e^{-i\omega t'}g(t'),
\eeq
due to the boundary conditions (\ref{bcg}). 
As in Sec.~\ref{two}, it is also possible to flatten the excitation energy curve versus $\omega$ around $\omega_0$ by 
increasing the multiplicity of the zero, i.e., simply choosing 
$\omega_1=\omega_2=...=\omega_N=\omega_0$.   
In Fig.~\ref{fig1}, we compare the one, two, and three frequency protocols with the choice $\omega_i=\omega_0$. We have plotted both the trajectory $x_0(t)$ and the transient excess of total energy $\Delta E (t)$ during the transport.  
The parameters chosen for Fig.~\ref{fig1} are inspired by Ref.~\cite{Walther} in which a transport of single $^{40}$Ca$^+$ ions is performed over a distance 20000 $a_0$  where $a_0=(\hbar/m\omega_0)^{1/2}$ is the harmonic length associated with the angular frequency $\omega_0=2\pi \times 1.41$ MHz, in a time $\omega_0t_f=2\pi \times 5$. We have chosen for Fig.~\ref{fig1} the same atom and angular frequency $\omega_0$ but have considered a transport over a larger distance, $d=30000\;a_0$, realized over a much shorter time duration $t_f$ such that $\omega_0t_f=2\pi \times 1.25$.

We clearly observe in Fig.~\ref{fig1b} an impressive increase of the robustness against the variations of $\omega$ about $\omega_0$ through the increasing local flatness about $\omega_0$ when $N$ increases. A price to pay to benefit from this robustness is a more involved trap trajectory with a clear non monotonous character (Fig.~\ref{fig1}a) and with an increasingly large transient energy (Fig.~\ref{fig1}b) \cite{footnote}. The oscillatory character of the trajectory can be intuitively understood. Indeed to ensure an optimal transport even for a trap frequency slightly smaller or larger than $\omega_0$, one has to design a trajectory that compensates for the delay or advance that the two types of trapping about $\omega_0$ will imply. Such strategies are reminiscent of the spin echo technique in which a succession of pulses is used to focus the spins towards the desired state even though they experience different Rabi frequencies 
\cite{Molmer}.

\begin{figure}[h]
\includegraphics[width=8cm]{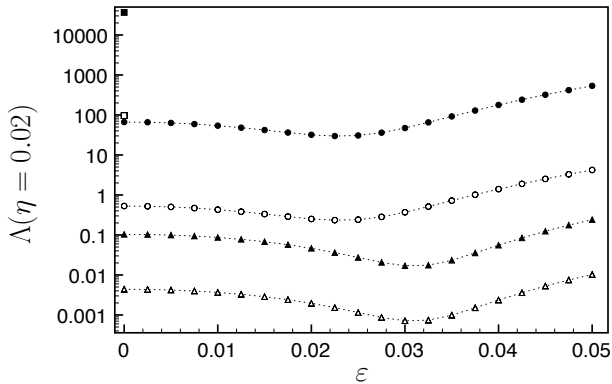}
\caption{Robustness function $\Lambda(\eta)$ for $\eta=0.02$ for three different protocols: one-point protocol with $\omega_1=\omega_0$ (square), two-point protocol with $\omega_1=\omega_0(1-\varepsilon)$ and $\omega_2=\omega_0(1+\varepsilon)$ (circle) and the three-point protocol with $\omega_1=\omega_0(1-\varepsilon)$, $\omega_2=\omega_0$,  and $\omega_3=\omega_0(1+\varepsilon)$ (triangle). The transport time is $\omega_0t_f=2\pi \times 1.25$ for the filled symbols, and $\omega_0t_f=2\pi \times 2.5$ for the open symbols.}
\label{fig2}
\end{figure}

The performance of the protocols can be evaluated by means of the function
\begin{equation}
\Lambda(\eta) = \frac{1}{2\omega_0\eta} \int_{\omega_0(1-\eta)}^{\omega_0(1+\eta)} \frac{\Delta E(t_f)}{\hbar \omega_0} {\rm d}\omega,
\end{equation}
which gives the average excitation number over a finite range of frequencies about the central angular frequency $\omega_0$. It therefore measures the robustness of the transport against the frequency of the trap. Figure \ref{fig2} compares the performance of the 1,2 and 3-point protocols for $\eta=0.02$ and for two different final times. The abscissa $\varepsilon$ measures the interval between chosen frequencies for 2 and 3 points, see the Figure caption. If not stated otherwise, the parameters for Fig.~\ref{fig2} are the same as for Fig.~\ref{fig1}.

The observed general trends are intuitive: for a given protocol increasing the final time improves the robustness, the $N$-point protocol yields an improvement by more than two orders of magnitude compared to the $(N-1)$-point protocol. For the shortest transport time, the optimal choice for the 3-point protocol improves the robustness of the transport by more than five orders of magnitude compared to the 1-point protocol. Figure~\ref{fig2} also demonstrates the importance of the choice of the trap frequencies $\omega_i$ and the optimal intervals. This optimization shows that the protocol that relies on the same frequencies $\omega_i=\omega_0$ is never the best strategy. For instance, the three-point protocol with $\omega_1=0.97 \omega_0$, $\omega_2=\omega_0$ and $\omega_3=1.03\omega_0$ reduces the average excitation energy by a factor of more than 5 compared to the 3-point protocol with $\omega_1=\omega_2=\omega_3=\omega_0$. 

%
%
%
%
%
\section{Discussion\label{dis}}
We have designed systematic fast transport protocols to leave the atoms (isolated, or in condensates) translationally unexcited 
at final time for a  range of trap frequencies. This is instrumental in avoiding the effect of instability of trap frequencies
among different runs of the experiment.  
Compared to a previous method (perturbative with respect to the spring constant deviations) 
to robustify the transport function 
\cite{Lu14}, the current approach is simpler to implement, contains no approximations, and gives  an explicit form for the transport function for a chosen trap-frequency domain of stability.    
The same methods put forward here may be applied to other systems as well.      
Consider in particular two particles 1 and 2 confined by two different harmonic potentials with angular frequency $\omega_1$ and $\omega_2$ respectively. For instance, two different atoms in the same dipole trap or two atoms of the same species but in different Zeeman state that are magnetically trapped. We assume that the two particles do not interact. To transport optimally such a dual species system, we need to guarantee a vanishing energy excess for both types of atoms. The results of the 2-point protocol developed in Sec.~\ref{two} can then be applied directly. Its robustness can also be improved using higher order protocols as explained in the previous section.
The technique to design optimal and robust trap trajectories can be readily generalized to more than two species following directly Sec.~\ref{urp}. Finally, the method described may also be applied to a broader set of physical inverse problems, 
e.g. in optics, to nullify the Fourier transform of a controllable function in a given interval.    
%
%
%
%
%
\acknowledgements{It is a pleasure to thank An Shuoming of Tsinghua University, and Xi Chen and Qi Zhang of Shanghai University for useful comments. This study has been partially supported through the grant NEXT  ANR-10-LABX-0037 in the framework of the Ç Programme des Investissements dÕAvenir and the Institut Universitaire de France. We also acknowledge funding by Basque Country Government (Grant No. IT472-10), Ministerio de Econom\'{i}a y Competitividad (Grant No.
FIS2012-36673-C03-01), and the program UFI 11/55.}
\appendix
\section{Exact solution}
To find the exact solution of Eq.~(\ref{eq2}) we search for a solution of the form \cite{Husimi,NIST06}
\begin{equation}
\Psi(x,t) = \Phi(X,t)e^{i\gamma X},
\label{eq3}
\end{equation}
where $\gamma$ is a time dependent variable,  and $X=x-x_c(t)$ is the position shifted by a scalar time dependent parameter $x_c(t)$ to be determined. In the following we show how $\gamma$ and $x_c$ are related self-consistently. For this purpose, we calculate separately the different terms of the time-dependent Schr\"odinger solution:
\begin{eqnarray}
\!\!\!\!\!\!\!\!&&\!\!\!\! \frac{\partial \Psi(x,t)}{\partial t} = \left[\! -\dot{x}_c\frac{\partial
\Phi}{\partial X}\!+\!\frac{\partial \Phi}{\partial t}\!+\!
i(\dot{\gamma}X\!-\!\gamma \dot{x}_c)\Phi
\right]\!\!e^{i\gamma X}\!,
\label{eqa1} 
\\
\!\!\!\!\!\!\!\!
&-&\frac{\hbar^2}{2m} \frac{\partial^2 \Psi}{\partial x^2}  = 
-\frac{\hbar^2}{2m}\! \left(\! \frac{\partial^2 \Phi}{\partial
X^2}+2\gamma i \frac{\partial \Phi}{\partial X}
-\gamma^2\Phi\! \right)\!\!e^{i\gamma X}\!,
\label{eqa2}\\
\!\!\!\!\!\!\!\!&&V(x,t) \Psi(x,t)  = \frac{1}{2}m\omega_0^2( X+x_c-x_0)^2\Phi e^{i\gamma X}. 
\label{eqa3}
\end{eqnarray}
Combining Eqs.~(\ref{eqa1}), (\ref{eqa2}) and (\ref{eqa3}), we obtain
\begin{eqnarray}
i\hbar \frac{\partial \Phi}{\partial t} & = & -\frac{\hbar^2}{2m}
\frac{\partial^2 \Phi}{\partial
X^2}+\frac{1}{2}m\omega_0^2X^2\Phi \nonumber \\
 & + &  i\hbar\left( \dot x_c - \frac{\hbar \gamma}{m}\right)\frac{\partial \Phi}{\partial
X}\nonumber \\
 & + & [ \hbar \dot{\gamma}+m\omega_0^2(x_c-x_0)]X\Phi \nonumber \\
 & + &\! \left[\! \hbar \gamma \dot{x}_c + \frac{\hbar^2}{2m}\gamma^2+\frac{1}{2}m\omega_0^2(x_c-x_0)^2\right]\!\Phi.
 \label{eqa4}
\end{eqnarray}
By setting to zero the factor of $\partial\Phi/\partial X$ and also the factor of $X\Phi$ we get the relation between $\gamma$ and $x_c$, $\gamma=m\dot x_c/m$ and the equation of motion of the variable $x_c$: $\ddot{x}_c+\omega_0^2(x_c-x_0)=0$. The variable $x_c$ corresponds to the trajectory of the classical counterpart of the quantum transport problem. The last term of Eq.~(\ref{eqa4}) is a scalar time dependent term which contributes as a time dependent phase. To remove it we introduce the wave function $\tilde \Phi$ defined by 
\begin{equation}
\Phi(X,t) = \tilde{\Phi}(X,t) \exp\left(
\frac{i}{\hbar}\int_0^tdt'{\cal L}(t')\right), \label{eq9}
\end{equation}
where ${\cal L}(t)$ is the classical-mechanical Lagrangian of the transport,
\begin{equation}
{\cal L}(t) =  \frac{m}{2}\dot{x}_c^2 -
\frac{1}{2}m\omega_0^2(x_c-x_0)^2.
\label{eq9b}
\end{equation}
With this choice, the wave function $\tilde \Phi$ obeys the time dependent Schr\"odinger equation for a \emph{static} harmonic potential of angular frequency $\omega_0$,
\begin{equation}
i\hbar \frac{\partial \tilde{\Phi}}{\partial t} =
-\frac{\hbar^2}{2m} \frac{\partial^2 \tilde{\Phi}}{\partial
X^2}+\frac{1}{2}m\omega_0^2X^2\tilde{\Phi}(X,t). \label{eq10}
\end{equation}
Finally, the exact expression for the solution of Eq.~(\ref{eq2}) takes the form
\begin{equation}
\Psi(x,t) = \tilde{\Phi}(X,t)e^{\displaystyle
\frac{imX\dot{x}_c}{\hbar}} e^{\displaystyle
\frac{i}{\hbar}\int_0^tdt'{\cal L}(t')}.
\label{eq11}
\end{equation}
\section{Energy}
The instantaneous energy reads $E(t) = \langle \Psi(t) | H_0(t) | \Psi(t) \rangle$. To perform this calculation using the solution (\ref{eq11}), it is convenient to write the potential in the form $V(x-x_0)=V(x-x_c + x_c-x_0)=V(X+x_c-x_0)$. As $V$ is quadratic there are three contributions to the energy:
\begin{eqnarray}
E(t) & = & \int dX \Psi^*(x,t)\left( -\frac{\hbar^2}{2m}\frac{d^2}{dX^2} + \frac{1}{2}m\omega_0^2X^2 \right)\Psi^*(x,t)
\nonumber \\
& +  & \frac{m}{2}\omega_0^2(x_c-x_0)^2
\nonumber \\
& + &  m\omega_0^2(x_c-x_0)\int dX \tilde\Phi^*(X,t)X\tilde\Phi(X,t).
\label{een}
\end{eqnarray}
Assuming that the initial state corresponds to the n-th eigenstate of the harmonic potential, we have $\tilde \Phi (X,t)=\varphi_n(X)e^{-iE_nt/\hbar}$ with $E_n=\hbar (n+1/2)\omega_0$ and the last integral of Eq.~(\ref{een}) vanishes by parity. The first term can be readily calculated. We thus find
\begin{equation}
E(t)=E(0) + \frac{m}{2}\dot x_c^2 + \frac{m}{2}\omega_0^2 (x_c-x_0)^2.
\end{equation}
Let us introduce the position $\xi=x_c-x_0$ of the fictitious classical particle in the frame of the moving potential. This position obeys the equation
\begin{equation}
\ddot \xi + \omega_0^2 \xi = - \ddot x_0.
\end{equation}
The solution of this equation provided that $x_c(0)=0$ and $\dot x_c(0)=0$ reads
\begin{equation}
\xi(t) = -\frac{1}{\omega_0} \int_0^t dt' \ddot x_0 \sin [\omega_0 (t-t')].
\end{equation}
Interestingly, we can write this solution in the complex form
\begin{equation}
a(t)=\xi -i\frac{\dot \xi}{\omega_0} = \frac{i}{\omega_0}e^{i\omega_0t}\int_0^t \ddot x_0(t')e^{-i\omega_0t'}dt'.
\end{equation}
The instantaneous energy can be also reexpressed in terms of $a$:
\begin{equation}
\Delta E(t)=E(t)-E(0)=m\dot \xi \dot x_0 + \frac{m}{2}\dot x_0^2 + \frac{m\omega_0^2}{2}|a(t)|^2.
\end{equation}
For the transport problem we are interested in \cite{NIST06}
\begin{eqnarray}
\Delta E(t_f) = \frac{m}{2}\bigg|\int_0^{t_f}\ddot x_0(t')e^{-i\omega_0 t'}dt'\bigg|^2. 
\label{eqdifapb}
\end{eqnarray}
In terms of dimensionless variables
for time and trap-position: 
\beqa
T&=&t\omega_0,
\nonumber\\
y_0&=&x_0/a_0,
\eeqa
where $a_0=[\hbar/(m\omega_0)]^{1/2}$ is the characteristic length for the harmonic oscillator, 
the excitation energy in units of the vibrational quantum $\hbar\omega_0$ takes the simple form
\beq
\frac{\Delta E(T_f)}{\hbar\omega_0}=\frac{1}{2}\left|\int_0^{T_f} {y_0''} e^{-iT'} dT'\right|^2
\eeq
where $T_f=\omega_0t_f$ and the double prime represents the second derivative with respect to $T$.

Equation (\ref{eqdifapb}) provides us with an analogy between transport and Fraunhofer diffraction in optics. Indeed, the excess of energy is proportional to the modulus of the Fourier transform of the acceleration profile at the angular frequency $\omega_0$. The acceleration profile therefore plays the role of an optical transmittance. According to this analogy, an optimal transport corresponds to a dark fringe in wave optics (zero intensity) \cite{David08}. 

\end{document}